\documentclass[prl,aps,amsmath,amssymb,letterpaper,twocolumn,natbib]{revtex4}

\pdfoutput=1
\usepackage{hyperref}
\usepackage{graphicx}

\newcommand{\en}{\epsilon_{\rm n}}
\newcommand{\ed}{\epsilon_{\rm d}}

\def\beq{\begin{equation}}
\def\eeq{\end{equation}}
\def\bea{\begin{eqnarray}}
\def\eea{\end{eqnarray}}

\def\kt{k_{\rm B}T}

\begin{document}

\title{Control of pathways and yields of protein crystallization through the interplay of nonspecific and specific attractions}

\author{Stephen Whitelam} 
\email{swhitelam@lbl.gov}
\affiliation{Molecular Foundry, Lawrence Berkeley National Laboratory, 1 Cyclotron Road, Berkeley, CA 94720, USA}

\begin{abstract}
We use computer simulation to study crystal-forming model proteins equipped with interactions that are both orientationally specific and nonspecific. Distinct dynamical pathways of crystal formation can be selected by tuning the strengths of these interactions. When the nonspecific interaction is strong, liquidlike clustering can precede crystallization; when it is weak, growth can proceed via ordered nuclei. Crystal yields are in certain parameter regimes enhanced by the nonspecific interaction, even though it promotes association without local crystalline order. Our results suggest that equipping nanoscale components with weak nonspecific interactions (such as depletion attractions) can alter both their dynamical pathway of assembly and optimize the yield of the resulting material. 
\end{abstract}
\maketitle


Controlling the crystallization of molecular and nanoscale systems remains a principal challenge of physics and chemistry. Controlling protein crystallization in particular is central to protein characterization, but despite advances in our understanding of protein phase behavior and association dynamics~\cite{muschol1997liquid,sear1999phase,vekilov2005two,galkin2000control,foffi2002phase,filobelo2005ssc,pan2005nucleation,wolde1997epc, sear2007nucleation,doye2007controlling,liu2007vapor,gšgelein2008simple,auer2008self,chen2008fractal,liu2009self} we lack a set of rules for rational production of protein crystals {\em in vitro}~\cite{slabinski2007challenge}. Some proteins crystallize {\em in vivo}. S (`surface')-layer proteins form functional crystalline lattices on the outsides of many bacteria and archaea, and were among the first protein structures used to organize nanomaterials in a `bottom-up' fashion~\cite{messner1992crystalline,sleytr1997baa}. The sbpA S-layer protein from the bacterium {\em Lysinibacillus sphaericus} forms a square crystalline lattice of tetramers, and has been shown to crystallize in a `nonclassical' fashion on supported lipid bilayers {\em in vitro}~\cite{deyoreo}: order emerges from dense amorphous clusters, rather than directly from crystalline nuclei. A similar dynamics is thought to operate during crystallization of the globular protein lysozyme~\cite{vekilov2005two,galkin2000control}.

Here we introduce a molecular model designed to study crystallization in the presence and absence of amorphous intermediates. The model is inspired by the crystallization of the sbp S-layer, but is designed to be simple enough to allow us to draw conclusions about control of crystallization pathways more generally. The model comprises monomers equipped with two types of interaction. The first consists of a directionally nonspecific attraction, designed to mimic the tendency of proteins to associate in a manner that does not uniquely constrain the orientations of neighboring monomers. The second interaction comprises directionally- and chemically specific attractive patches whose placement is suggested by the S-layer's electron density map~\cite{norville20077Œ} and its unusual crystal structure. Patches predispose monomers to the formation of a square crystalline lattice of tetramers. Here we attempt to answer the following question: How does the nonspecific interaction influence the dynamics of formation and yields of crystals whose symmetries are selected by the specific attraction?
 
In what follows we show that distinct dynamical pathways of crystal formation can be selected by tuning the strengths of nonspecific and specific interactions (this selection is suggested by the bulk free energy landscape of generic anisotropic particles~\cite{whitelam2010non}, and by distinct dynamical pathways seen in simulation studies of virus capsid assembly~\cite{wilber2007reversible} and polymer crystallization~\cite{hu2005polymer}). Nonclassical assembly via liquidlike intermediates is possible when the nonspecific interaction is strong; when it is weak, classical modes of assembly can be realized. In the former regime the lifetime of the liquidlike phase can be controlled by varying the strength of the specific interaction. We show also that optimal crystallization conditions are found when the nonspecific interaction is nonzero -- a result striking in light of the fact that this interaction promotes none of the symmetries of the crystal -- but not strong enough to induce the formation of large liquidlike intermediates. Other model proteins bearing both nondirectional and directional attractions have recently been studied, yielding valuable insight into phase behaviors and crystallization dynamics as temperature is varied~\cite{gšgelein2008simple,liu2009self}. The present study is distinguished by its exploration of the dynamics and yields of assembly for nonspecific and specific interactions of varying absolute {\em and} relative strength. Such an exploration is required in order to assess monomers' possible modes of assembly. 

\begin{figure} [h!]
\label{}
\centering
\includegraphics[width=\linewidth]{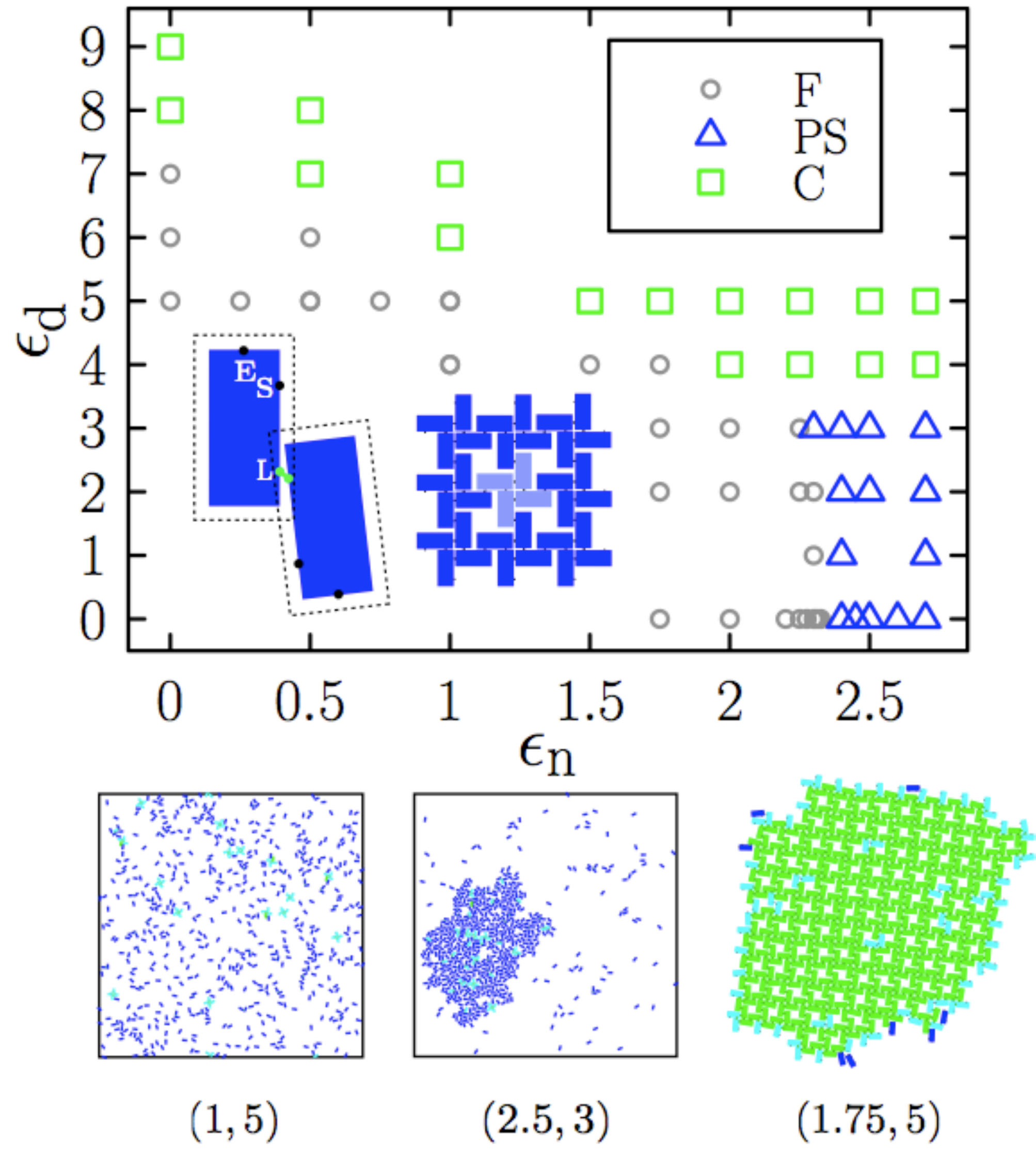} 
\caption{\label{fig1} Model geometry and phase diagram. Inset: Monomers consist of rectangles equipped with an attractive rectangular force-field (dotted) and decorated with three sticky patches labeled E, S and L. Only L-L and E-S pairings are reactive. Patch geometry predisposes monomers to the formation of a square lattice of tetramers (sketched), in mimicry of the sbpA S-layer. Main figure: Model phase diagram in the space of specific $\ed$ and nonspecific $\en$ interaction strengths (600 particles, 10.91\% coverage by area) shows regimes of homogeneous fluid (F), phase-separated liquid and vapor (PS), and crystal order (C). Snapshots below show examples of phases F, PS, and C, from left to right, taken from points $(\epsilon_{\rm n}, \epsilon_{\rm d})$.}
 \end{figure}

Model geometry is shown in Fig.~\ref{fig1}. The model comprises a featureless two-dimensional substrate on which live, in continuous space, hard rectangular monomers of small edge length $a$ and aspect ratio $2.2$. Monomers possess both specific and nonspecific pairwise interactions. Specific interactions are mediated by three sticky patches placed on two sides of the rectangle, as shown, each a distance $a/2$ from the nearest vertex. Patches are of type E (`edge'), S (`short-arm') and L (`long-arm'), and are selectively reactive: a `directional' bond of energy $-\epsilon_{\rm d} \kt$ is made when two L patches or one E- and one S patch are separated by a distance of less than $a/5$. Patch geometry predisposes monomers to the formation of a square lattice of tetramers, as sketched, in mimicry of the sbpA S-layer~\cite{norville20077Œ}. The tetrameric repeat unit of the latter measures about 18 nm on a side, and for correspondence we imagine $a \approx 4$ nm. In the simulations discussed below we defined particles making two directional bonds to be `partially crystalline' (rendered light blue in snapshots), and particles making three directional bonds to be `crystalline' (rendered green in snapshots). We denote by $f_{\rm p}$ and $f_{\rm c}$ the fractions of monomers in partially crystalline and crystalline states, respectively. The nonspecific interaction is a pairwise bond of energy $-\epsilon_{\rm n} \kt$, and is activated by the overlap of the two dotted rectangles shown; these rectangles are concentric with the monomers that give rise to them, and have side lengths $2a/5$ in excess of the sides of those monomers. Interaction ranges assume solution conditions to be such that protein-protein interactions are attenuated on a length scale of about 1 nm.

We performed two types of $NVT$ simulations within periodically-replicated, square boxes: `sampling' and `dynamic'. Sampling simulations (designed to probe thermal equilibrium) employed 600 particles whose total area comprised 10.91\% of the simulation box. Simulations were started from a configuration consisting of a square crystal (or a cluster of noncrystalline tetratic order) inserted into a vapor of monomers. We propagated these systems using local Monte Carlo moves supplemented by the nonlocal `teleportation' algorithm described in the supplementary material~\footnote{Linked from this article's arXiv page.}. Dynamic simulations were begun from configurations of randomly dispersed and oriented monomers, and were propagated using a `virtual-move' Monte Carlo algorithm~\cite{whitelam2007auk,whitelam2009rcm}. Its purpose is to approximate a diffusive dynamics by using potential energy gradients to effect collective translations and rotations ignored by standard single-particle algorithms. Accounting for collective modes of motion is necessary in order to identify when a molecular system undergoing overdamped motion might assembly robustly or become kinetically trapped. We performed simulations of either 600 or 2000 monomers, and considered monomer occupancies by area ranging from 20\% to 1\% (focusing on the case of 10.91\%). Further details of simulation protocol (and phase classifications) can be found in the supplementary material.

The phase diagram for the model in the $(\epsilon_{\rm n}, \epsilon_{\rm d})$ plane, derived from sampling simulations, is shown in Fig.~\ref{fig1}. It identifies regimes of homogeneous fluid, phase-separated liquid and vapor, and crystalline order. The structure of the phase diagram is similar to that computed by mean field theory applied to prototypical anisotropic particles~\cite{whitelam2010non}: notably, liquid-vapor phase separation is in large part driven by the nonspecific interaction, and moderate values of this interaction enlarge the regime of crystal stability. For larger values of $\epsilon_{\rm n}$ we observe the emergence of a noncrystalline tetratic phase that owes its existence to monomers' rectangular shape~\cite{geng2009theory,donev2006tetratic} (see Fig. S1). Association driven by the nonspecific interaction stabilizes none of the order characteristic of the crystal: we find that $\langle f_{\rm c} \rangle = 0$ when $\epsilon_{\rm d}=0$ (Fig. S1).

\begin{figure}[ht]
\label{}
\centering
\includegraphics[width=\linewidth]{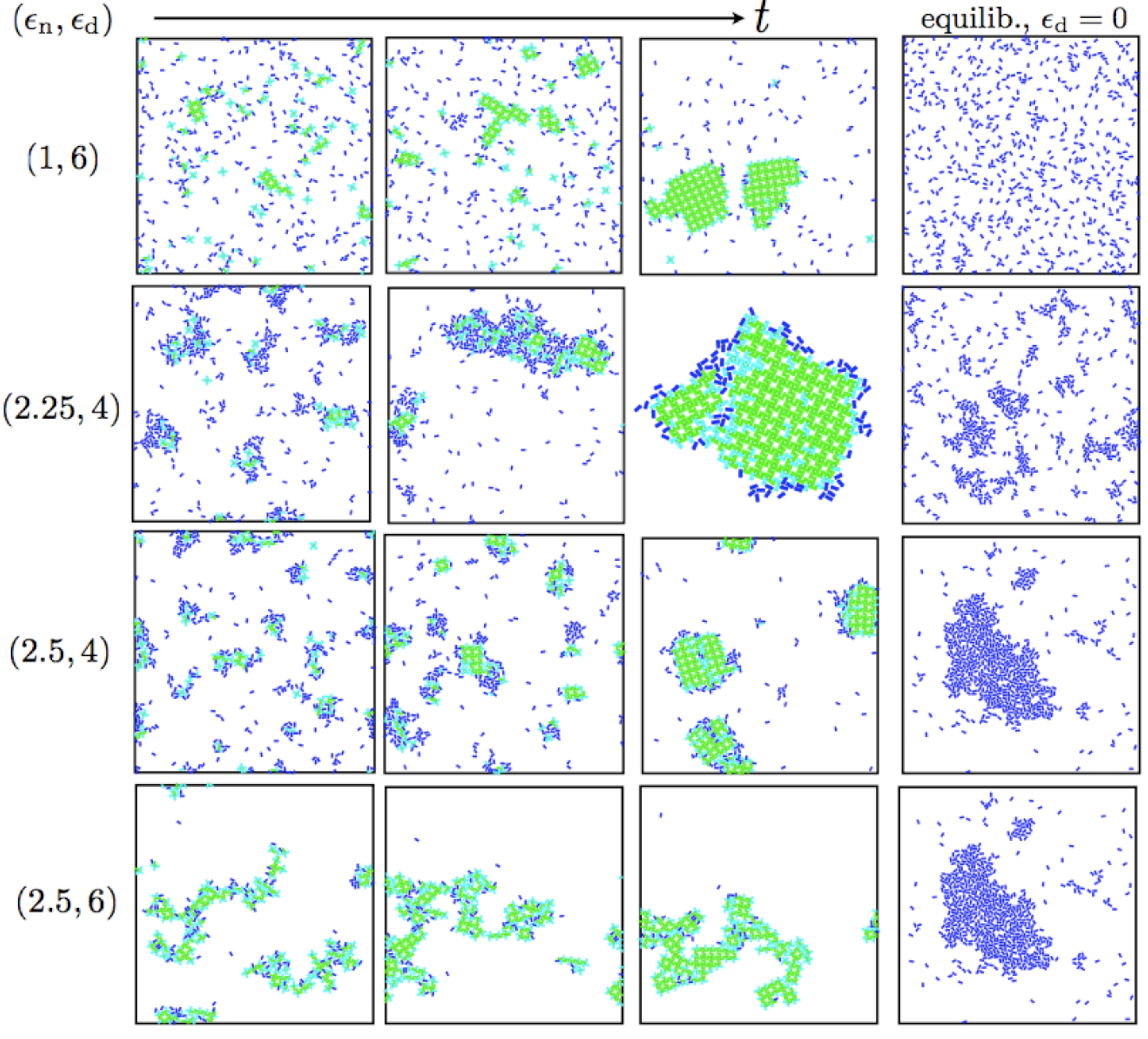} 
\caption{\label{fig2} 
Time-ordered snapshots from dynamic simulations (600 monomers, 10.91\% coverage by area) for four different choices of $(\en,\ed)$. Mechanisms of crystal assembly range from classical (top row), where the growing nucleus possesses the architecture of the stable solid, to nonclassical (rows 2 and 3), where the crystal emerges from the midst of dense liquidlike clusters. There also exist regimes (e.g. bottom row) in which the formation of gel-like networks prevents the emergence of the metastable liquid phase. At right: typical configurations in equilibrium in the absence of the specific interaction.}
 \end{figure}

We next used dynamical simulation to determine how crystals form in different regions of parameter space. We found that crystallization can proceed by dynamical pathways both nonclassical -- along which metastable liquidlike precursors form and only subsequently acquire crystalline order -- and classical, along which the critical nucleus possesses the architecture of the stable solid. In Fig.~\ref{fig2} we show examples of both pathways. In general, the greater the value of $\en$ the greater the propensity for liquidlike clustering to precede crystallization (in a related vein, liquidlike clusters can precede the formation of model virus capsids if interaction patch specificities are sufficiently low~\cite{wilber2007reversible}). If $\en$ is large enough to induce liquid-vapor phase separation then the resulting dynamics can comprise crystallization-arrested spinodal decomposition. This dynamics resembles that seen in experiment~\cite{deyoreo}. However, within this regime increasing $\ed$ can shorten the lifetime of the metastable liquid by enhancing crystallization kinetics or (for large enough $\ed$) inducing assembly of gel-like intermediates (Fig.~\ref{fig2} and Fig. S2). This result is reminiscent of the kinetic stabilization of amorphous phases seen in computer simulations of liquid crystals~\cite{henrich2010ordering}, and suggests that in our model, as in that work, there exist regions of phase space within which Ostwald's step rule does not hold. The latter states that the liquid phase, if stable with respect to the homogeneous fluid and metastable with respect to the crystal, should emerge prior to crystallization.

\begin{figure*} 
\label{}
\centering
\includegraphics[width=0.9\linewidth]{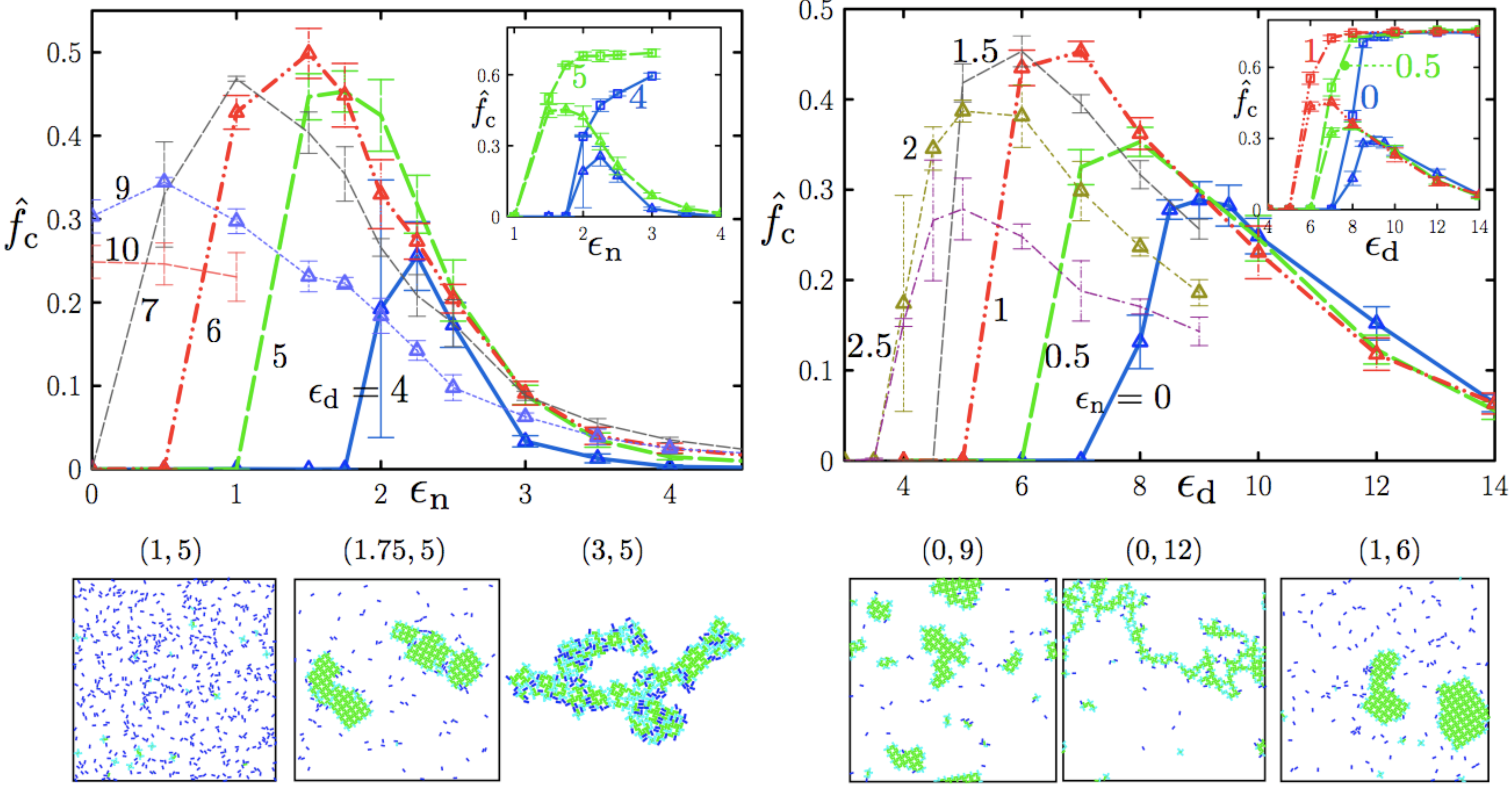} 
\caption{\label{fig3} Long-time scaled yields $\hat{f}_{\rm c}$ from dynamic simulations at specified fixed values of $(\en,\ed)$ (600 particles, 10.91\% coverage by area; values of $\ed$ and $\en$ label lines in left and right panels, respectively). Data points represent the mean of 5 independent simulations; lines are a guide to the eye. Insets compare selected sets of dynamic simulations with their equilibrium counterparts. The left panel shows the general enhancement of crystal yield conferred by nonzero $\en$, for given $\ed$ (up to $\ed \approx 9$). The right panel shows that `best' assembly is found in general for nonzero $\en$, even though the nonspecific interaction stabilizes none of the symmetries of the crystal. Snapshots below show configurations from simulations at specified $(\en,\ed).$}
 \end{figure*}

To determine the effect of the nonspecific interaction upon crystal quality we measured scaled yields $\hat{f}_{\rm c} \equiv f_{\rm c} \left(f_{\rm c}/\left( f_{\rm p}+f_{\rm c}\right) \right)^2$ (an order parameter that rewards compact crystals with a large bulk-to-surface ratio) per particle after long dynamic simulations (Fig. S3) for fixed values of $(\en,\ed)$. The left panel of Fig.~\ref{fig3} illustrates the effect of increasing $\en$ given $\ed$. For given $\ed \, (\lesssim 9)$, small values of $\en$ enhance assembly of the crystal, while large values induce dynamic arrest (cf. equilibrium behavior (inset); see also Fig. S4). The value of $\en$ at which arrest occurs is a function of $\ed$: in general, optimal assembly for given $\ed$ occurs when $\en$ is too small to induce the formation of large liquidlike clusters, in accord with a suggestion made on the basis of a study of isotropic model proteins~\cite{chen2008fractal}. For certain choices of the specific interaction, however, such as $\ed=4$, yields are maximized close to the liquid-vapor critical region. For larger $\ed$ $(> 9)$, nonzero $\en$ provides little or no enhancement of yield. The right panel further reveals that the regime of best assembly occurs for small but nonzero values of $\en$; we observed similar behavior at monomer concentrations of 1\% by area (Fig. S5). This enhancement of yield by the nonspecific interaction is striking in light of the fact that the latter promotes association without stabilizing the local order of the crystal. This result evokes one obtained from simulation studies of the self-assembly of closed virus capsids, namely, that capsid yield is optimized by interaction patch specificities that are neither too high nor too low~\cite{hagan2006dynamic,wilber2007reversible}. We speculate that in our model this enhancement has the following origin. Partial reversibility -- the ability of components to transiently break bonds in order to correct the nascent structural defects of growing assemblies -- is a necessary condition for robust self-assembly~\cite{whitesides2002bms,break2,new_rapaport}. Particles bearing moderately strong nonspecific and specific interactions and particles equipped with very strong specific interactions may form solids of similar thermodynamic stability. However, is likely that the former gives rise to a greater degree of `partial reversibility' than does the latter: it is easier to break in sequence two moderately strong bonds than one very strong bond in order to correct nascent defects as structures grow. It is likely also that at very low monomer concentrations the increased collisional cross-section associated with the nonspecific interaction leads to an enhanced kinetics of assembly.

We have used computer simulation to study a model of crystal-forming monomers equipped with interactions that are both nonspecific, in an orientational and chemical sense, and specific. Distinct dynamical pathways of crystal formation can be selected by tuning the strengths of these interactions. Fluctuations of density and structure sometimes cooperate (enhancing assembly), and sometimes conflict (impairing assembly). While both scenarios are suggested by simulations of isotropic model proteins~\cite{wolde1997epc,chen2008fractal}, here the presence of two types of interaction allows such fluctuations to be varied in strength (at fixed temperature and concentration) with a high degree of independence. We do not know the extent to which the qualitative findings of our solvent-free, two-dimensional simulations are relevant to real nanoscale components in three dimensions -- where, for instance, reorganization of liquidlike intermediates might be considerably more rapid than in 2d -- but direct extrapolation suggests that trading specific- for nonspecific interaction strength can alter assembly pathways and might be one way to optimize assembly. In protein solutions one could change the respective magnitudes of specific and nonspecific interactions by altering ionic strength and using inert nanoscale components to induce a depletion attraction~\cite{marenduzzo2006depletion}. Our results suggest that by trading specific- for nonspecific interaction strength, proteins with similar values of the second virial coefficient $B_2$ can be made to crystallize in dynamically distinct ways and with different propensities (Fig. S6). This suggestion is consistent with the observation~\cite{george19976} that even proteins possessing values of $B_2$ within the `crystallization slot' are not guaranteed to crystallize. 


We thank Sungwook Chung, Seong-Ho Shin, Jim DeYoreo, Carolyn Bertozzi and Caroline Ajo-Franklin for discussions. This work was performed at the Molecular Foundry, Lawrence Berkeley National Laboratory, and was supported by the Director, Office of Science, Office of Basic Energy Sciences, of the U.S. Department of Energy under Contract No. DE-AC02--05CH11231.


\begin{thebibliography}{34}
\expandafter\ifx\csname natexlab\endcsname\relax\def\natexlab#1{#1}\fi
\expandafter\ifx\csname bibnamefont\endcsname\relax
  \def\bibnamefont#1{#1}\fi
\expandafter\ifx\csname bibfnamefont\endcsname\relax
  \def\bibfnamefont#1{#1}\fi
\expandafter\ifx\csname citenamefont\endcsname\relax
  \def\citenamefont#1{#1}\fi
\expandafter\ifx\csname url\endcsname\relax
  \def\url#1{\texttt{#1}}\fi
\expandafter\ifx\csname urlprefix\endcsname\relax\def\urlprefix{URL }\fi
\providecommand{\bibinfo}[2]{#2}
\providecommand{\eprint}[2][]{\url{#2}}

\bibitem[{\citenamefont{Muschol and Rosenberger}(1997)}]{muschol1997liquid}
\bibinfo{author}{\bibfnamefont{M.}~\bibnamefont{Muschol}} \bibnamefont{and}
  \bibinfo{author}{\bibfnamefont{F.}~\bibnamefont{Rosenberger}},
  \bibinfo{journal}{J. Chem. Phys.} \textbf{\bibinfo{volume}{107}},
  \bibinfo{pages}{1953} (\bibinfo{year}{1997}).

\bibitem[{\citenamefont{Sear}(1999)}]{sear1999phase}
\bibinfo{author}{\bibfnamefont{R.~P.} \bibnamefont{Sear}}, \bibinfo{journal}{J.
  Chem. Phys.} \textbf{\bibinfo{volume}{111}}, \bibinfo{pages}{4800}
  (\bibinfo{year}{1999}).

\bibitem[{\citenamefont{Vekilov}(2005)}]{vekilov2005two}
\bibinfo{author}{\bibfnamefont{P.~G.} \bibnamefont{Vekilov}},
  \bibinfo{journal}{J. Crystal Growth} \textbf{\bibinfo{volume}{275}},
  \bibinfo{pages}{65} (\bibinfo{year}{2005}).

\bibitem[{\citenamefont{Galkin and Vekilov}(2000)}]{galkin2000control}
\bibinfo{author}{\bibfnamefont{O.}~\bibnamefont{Galkin}} \bibnamefont{and}
  \bibinfo{author}{\bibfnamefont{P.~G.} \bibnamefont{Vekilov}},
  \bibinfo{journal}{Proc. Nat. Acad. Sci.} \textbf{\bibinfo{volume}{97}},
  \bibinfo{pages}{6277} (\bibinfo{year}{2000}).

\bibitem[{\citenamefont{Foffi et~al.}(2002)\citenamefont{Foffi, McCullagh,
  Lawlor, Zaccarelli, Dawson, Sciortino, Tartaglia, Pini, and
  Stell}}]{foffi2002phase}
\bibinfo{author}{\bibfnamefont{G.}~\bibnamefont{Foffi}},
  \bibinfo{author}{\bibfnamefont{G.~D.} \bibnamefont{McCullagh}},
  \bibinfo{author}{\bibfnamefont{A.}~\bibnamefont{Lawlor}},
  \bibinfo{author}{\bibfnamefont{E.}~\bibnamefont{Zaccarelli}},
  \bibinfo{author}{\bibfnamefont{K.~A.} \bibnamefont{Dawson}},
  \bibinfo{author}{\bibfnamefont{F.}~\bibnamefont{Sciortino}},
  \bibinfo{author}{\bibfnamefont{P.}~\bibnamefont{Tartaglia}},
  \bibinfo{author}{\bibfnamefont{D.}~\bibnamefont{Pini}}, \bibnamefont{and}
  \bibinfo{author}{\bibfnamefont{G.}~\bibnamefont{Stell}},
  \bibinfo{journal}{Phys. Rev. E} \textbf{\bibinfo{volume}{65}},
  \bibinfo{pages}{31407} (\bibinfo{year}{2002}).

\bibitem[{\citenamefont{Filobelo et~al.}(2005)\citenamefont{Filobelo, Galkin,
  and Vekilov}}]{filobelo2005ssc}
\bibinfo{author}{\bibfnamefont{L.~F.} \bibnamefont{Filobelo}},
  \bibinfo{author}{\bibfnamefont{O.}~\bibnamefont{Galkin}}, \bibnamefont{and}
  \bibinfo{author}{\bibfnamefont{P.~G.} \bibnamefont{Vekilov}},
  \bibinfo{journal}{J. Chem. Phys.} \textbf{\bibinfo{volume}{123}},
  \bibinfo{pages}{014904} (\bibinfo{year}{2005}).

\bibitem[{\citenamefont{Pan et~al.}(2005)\citenamefont{Pan, Kolomeisky, and
  Vekilov}}]{pan2005nucleation}
\bibinfo{author}{\bibfnamefont{W.}~\bibnamefont{Pan}},
  \bibinfo{author}{\bibfnamefont{A.~B.} \bibnamefont{Kolomeisky}},
  \bibnamefont{and} \bibinfo{author}{\bibfnamefont{P.~G.}
  \bibnamefont{Vekilov}}, \bibinfo{journal}{J. Chem. Phys.}
  \textbf{\bibinfo{volume}{122}}, \bibinfo{pages}{174905}
  (\bibinfo{year}{2005}).

\bibitem[{\citenamefont{ten Wolde and Frenkel}(1997)}]{wolde1997epc}
\bibinfo{author}{\bibfnamefont{P.~R.} \bibnamefont{ten Wolde}}
  \bibnamefont{and} \bibinfo{author}{\bibfnamefont{D.}~\bibnamefont{Frenkel}},
  \bibinfo{journal}{Science} \textbf{\bibinfo{volume}{277}},
  \bibinfo{pages}{1975} (\bibinfo{year}{1997}).

\bibitem[{\citenamefont{Sear}(2007)}]{sear2007nucleation}
\bibinfo{author}{\bibfnamefont{R.~P.} \bibnamefont{Sear}}, \bibinfo{journal}{J.
  Phys. Cond. Mat.} \textbf{\bibinfo{volume}{19}}, \bibinfo{pages}{033101}
  (\bibinfo{year}{2007}).

\bibitem[{\citenamefont{Doye et~al.}(2007)\citenamefont{Doye, Louis, Lin,
  Allen, Noya, Wilber, Kok, and Lyus}}]{doye2007controlling}
\bibinfo{author}{\bibfnamefont{J.~P.~K.} \bibnamefont{Doye}},
  \bibinfo{author}{\bibfnamefont{A.~A.} \bibnamefont{Louis}},
  \bibinfo{author}{\bibfnamefont{I.~C.} \bibnamefont{Lin}},
  \bibinfo{author}{\bibfnamefont{L.~R.} \bibnamefont{Allen}},
  \bibinfo{author}{\bibfnamefont{E.~G.} \bibnamefont{Noya}},
  \bibinfo{author}{\bibfnamefont{A.~W.} \bibnamefont{Wilber}},
  \bibinfo{author}{\bibfnamefont{H.~C.} \bibnamefont{Kok}}, \bibnamefont{and}
  \bibinfo{author}{\bibfnamefont{R.}~\bibnamefont{Lyus}},
  \bibinfo{journal}{Phys. Chem. Chem. Phys} \textbf{\bibinfo{volume}{9}},
  \bibinfo{pages}{2197} (\bibinfo{year}{2007}).

\bibitem[{\citenamefont{Liu et~al.}(2007)\citenamefont{Liu, Kumar, and
  Sciortino}}]{liu2007vapor}
\bibinfo{author}{\bibfnamefont{H.}~\bibnamefont{Liu}},
  \bibinfo{author}{\bibfnamefont{S.~K.} \bibnamefont{Kumar}}, \bibnamefont{and}
  \bibinfo{author}{\bibfnamefont{F.}~\bibnamefont{Sciortino}},
  \bibinfo{journal}{J. Chem. Phys.} \textbf{\bibinfo{volume}{127}},
  \bibinfo{pages}{084902} (\bibinfo{year}{2007}).

\bibitem[{\citenamefont{G{\"o}gelein et~al.}(2008)\citenamefont{G{\"o}gelein,
  N{\"a}gele, Tuinier, Gibaud, Stradner, and
  Schurtenberger}}]{gšgelein2008simple}
\bibinfo{author}{\bibfnamefont{C.}~\bibnamefont{G{\"o}gelein}},
  \bibinfo{author}{\bibfnamefont{G.}~\bibnamefont{N{\"a}gele}},
  \bibinfo{author}{\bibfnamefont{R.}~\bibnamefont{Tuinier}},
  \bibinfo{author}{\bibfnamefont{T.}~\bibnamefont{Gibaud}},
  \bibinfo{author}{\bibfnamefont{A.}~\bibnamefont{Stradner}}, \bibnamefont{and}
  \bibinfo{author}{\bibfnamefont{P.}~\bibnamefont{Schurtenberger}},
  \bibinfo{journal}{J. Chem. Phys.} \textbf{\bibinfo{volume}{129}},
  \bibinfo{pages}{085102} (\bibinfo{year}{2008}).

\bibitem[{\citenamefont{Auer et~al.}(2008)\citenamefont{Auer, Dobson,
  Vendruscolo, and Maritan}}]{auer2008self}
\bibinfo{author}{\bibfnamefont{S.}~\bibnamefont{Auer}},
  \bibinfo{author}{\bibfnamefont{C.~M.} \bibnamefont{Dobson}},
  \bibinfo{author}{\bibfnamefont{M.}~\bibnamefont{Vendruscolo}},
  \bibnamefont{and} \bibinfo{author}{\bibfnamefont{A.}~\bibnamefont{Maritan}},
  \bibinfo{journal}{Phys. Rev. Lett.} \textbf{\bibinfo{volume}{101}},
  \bibinfo{pages}{258101} (\bibinfo{year}{2008}).

\bibitem[{\citenamefont{Chen et~al.}(2008)\citenamefont{Chen, Nellas, and
  Keasler}}]{chen2008fractal}
\bibinfo{author}{\bibfnamefont{B.}~\bibnamefont{Chen}},
  \bibinfo{author}{\bibfnamefont{R.~B.} \bibnamefont{Nellas}},
  \bibnamefont{and} \bibinfo{author}{\bibfnamefont{S.~J.}
  \bibnamefont{Keasler}}, \bibinfo{journal}{J. Phys. Chem. B}
  \textbf{\bibinfo{volume}{112}}, \bibinfo{pages}{4725} (\bibinfo{year}{2008}).

\bibitem[{\citenamefont{Liu et~al.}(2009)\citenamefont{Liu, Kumar, and
  Douglas}}]{liu2009self}
\bibinfo{author}{\bibfnamefont{H.}~\bibnamefont{Liu}},
  \bibinfo{author}{\bibfnamefont{S.~K.} \bibnamefont{Kumar}}, \bibnamefont{and}
  \bibinfo{author}{\bibfnamefont{J.~F.} \bibnamefont{Douglas}},
  \bibinfo{journal}{Phys. Rev. Lett.} \textbf{\bibinfo{volume}{103}},
  \bibinfo{pages}{18101} (\bibinfo{year}{2009}).

\bibitem[{\citenamefont{Slabinski et~al.}(2007)\citenamefont{Slabinski,
  Jaroszewski, Rodrigues, Rychlewski, Wilson, Lesley, and
  Godzik}}]{slabinski2007challenge}
\bibinfo{author}{\bibfnamefont{L.}~\bibnamefont{Slabinski}},
  \bibinfo{author}{\bibfnamefont{L.}~\bibnamefont{Jaroszewski}},
  \bibinfo{author}{\bibfnamefont{A.~P.~C.} \bibnamefont{Rodrigues}},
  \bibinfo{author}{\bibfnamefont{L.}~\bibnamefont{Rychlewski}},
  \bibinfo{author}{\bibfnamefont{I.~A.} \bibnamefont{Wilson}},
  \bibinfo{author}{\bibfnamefont{S.~A.} \bibnamefont{Lesley}},
  \bibnamefont{and} \bibinfo{author}{\bibfnamefont{A.}~\bibnamefont{Godzik}},
  \bibinfo{journal}{Protein Science} \textbf{\bibinfo{volume}{16}},
  \bibinfo{pages}{2472} (\bibinfo{year}{2007}).

\bibitem[{\citenamefont{Messner and Sleytr}(1992)}]{messner1992crystalline}
\bibinfo{author}{\bibfnamefont{P.}~\bibnamefont{Messner}} \bibnamefont{and}
  \bibinfo{author}{\bibfnamefont{U.}~\bibnamefont{Sleytr}},
  \bibinfo{journal}{Adv. Microbial Physiology} \textbf{\bibinfo{volume}{33}},
  \bibinfo{pages}{213} (\bibinfo{year}{1992}).

\bibitem[{\citenamefont{Sleytr}(1997)}]{sleytr1997baa}
\bibinfo{author}{\bibfnamefont{U.~B.} \bibnamefont{Sleytr}},
  \bibinfo{journal}{FEMS Microbiology Reviews} \textbf{\bibinfo{volume}{20}},
  \bibinfo{pages}{5} (\bibinfo{year}{1997}).

\bibitem[{\citenamefont{Chung et~al.}(2010)\citenamefont{Chung, Shin, Bertozzi,
  and De~Yoreo}}]{deyoreo}
\bibinfo{author}{\bibfnamefont{S.}~\bibnamefont{Chung}},
  \bibinfo{author}{\bibfnamefont{S.~H.} \bibnamefont{Shin}},
  \bibinfo{author}{\bibfnamefont{C.}~\bibnamefont{Bertozzi}}, \bibnamefont{and}
  \bibinfo{author}{\bibfnamefont{J.}~\bibnamefont{De~Yoreo}},
  \bibinfo{journal}{Submitted}  (\bibinfo{year}{2010}).

\bibitem[{\citenamefont{Norville et~al.}(2007)\citenamefont{Norville, Kelly,
  Knight, Belcher, and Walz}}]{norville20077Œ}
\bibinfo{author}{\bibfnamefont{J.~E.} \bibnamefont{Norville}},
  \bibinfo{author}{\bibfnamefont{D.~F.} \bibnamefont{Kelly}},
  \bibinfo{author}{\bibfnamefont{T.~F.} \bibnamefont{Knight}},
  \bibinfo{author}{\bibfnamefont{A.~M.} \bibnamefont{Belcher}},
  \bibnamefont{and} \bibinfo{author}{\bibfnamefont{T.}~\bibnamefont{Walz}},
  \bibinfo{journal}{J. Struct. Biol.} \textbf{\bibinfo{volume}{160}},
  \bibinfo{pages}{313} (\bibinfo{year}{2007}).

\bibitem[{\citenamefont{Whitelam}(2010)}]{whitelam2010non}
\bibinfo{author}{\bibfnamefont{S.}~\bibnamefont{Whitelam}},
  \bibinfo{journal}{J. Chem. Phys.} \textbf{\bibinfo{volume}{132}},
  \bibinfo{pages}{194901} (\bibinfo{year}{2010}).

\bibitem[{\citenamefont{Wilber et~al.}(2007)\citenamefont{Wilber, Doye, Louis,
  Noya, Miller, and Wong}}]{wilber2007reversible}
\bibinfo{author}{\bibfnamefont{A.~W.} \bibnamefont{Wilber}},
  \bibinfo{author}{\bibfnamefont{J.~P.~K.} \bibnamefont{Doye}},
  \bibinfo{author}{\bibfnamefont{A.~A.} \bibnamefont{Louis}},
  \bibinfo{author}{\bibfnamefont{E.~G.} \bibnamefont{Noya}},
  \bibinfo{author}{\bibfnamefont{M.~A.} \bibnamefont{Miller}},
  \bibnamefont{and} \bibinfo{author}{\bibfnamefont{P.}~\bibnamefont{Wong}},
  \bibinfo{journal}{J. Chem. Phys.} \textbf{\bibinfo{volume}{127}},
  \bibinfo{pages}{085106} (\bibinfo{year}{2007}).

\bibitem[{\citenamefont{Hu and Frenkel}(2005)}]{hu2005polymer}
\bibinfo{author}{\bibfnamefont{W.}~\bibnamefont{Hu}} \bibnamefont{and}
  \bibinfo{author}{\bibfnamefont{D.}~\bibnamefont{Frenkel}},
  \bibinfo{journal}{Interphases and Mesophases in Polymer Crystallization III}
  pp. \bibinfo{pages}{1--35} (\bibinfo{year}{2005}).

\bibitem[{\citenamefont{Whitelam and Geissler}(2007)}]{whitelam2007auk}
\bibinfo{author}{\bibfnamefont{S.}~\bibnamefont{Whitelam}} \bibnamefont{and}
  \bibinfo{author}{\bibfnamefont{P.~L.} \bibnamefont{Geissler}},
  \bibinfo{journal}{J. Chem. Phys.} \textbf{\bibinfo{volume}{127}},
  \bibinfo{pages}{154101} (\bibinfo{year}{2007}).

\bibitem[{\citenamefont{Whitelam et~al.}(2009)\citenamefont{Whitelam, Feng,
  Hagan, and Geissler}}]{whitelam2009rcm}
\bibinfo{author}{\bibfnamefont{S.}~\bibnamefont{Whitelam}},
  \bibinfo{author}{\bibfnamefont{E.~H.} \bibnamefont{Feng}},
  \bibinfo{author}{\bibfnamefont{M.~F.} \bibnamefont{Hagan}}, \bibnamefont{and}
  \bibinfo{author}{\bibfnamefont{P.~L.} \bibnamefont{Geissler}},
  \bibinfo{journal}{Soft Matter} \textbf{\bibinfo{volume}{5}},
  \bibinfo{pages}{1251} (\bibinfo{year}{2009}).

\bibitem[{\citenamefont{Geng and Selinger}(2009)}]{geng2009theory}
\bibinfo{author}{\bibfnamefont{J.}~\bibnamefont{Geng}} \bibnamefont{and}
  \bibinfo{author}{\bibfnamefont{J.~V.} \bibnamefont{Selinger}},
  \bibinfo{journal}{Phys. Rev. E} \textbf{\bibinfo{volume}{80}},
  \bibinfo{pages}{11707} (\bibinfo{year}{2009}).

\bibitem[{\citenamefont{Donev et~al.}(2006)\citenamefont{Donev, Burton,
  Stillinger, and Torquato}}]{donev2006tetratic}
\bibinfo{author}{\bibfnamefont{A.}~\bibnamefont{Donev}},
  \bibinfo{author}{\bibfnamefont{J.}~\bibnamefont{Burton}},
  \bibinfo{author}{\bibfnamefont{F.~H.} \bibnamefont{Stillinger}},
  \bibnamefont{and} \bibinfo{author}{\bibfnamefont{S.}~\bibnamefont{Torquato}},
  \bibinfo{journal}{Phys. Rev. B} \textbf{\bibinfo{volume}{73}},
  \bibinfo{pages}{54109} (\bibinfo{year}{2006}).

\bibitem[{\citenamefont{Henrich et~al.}(2010)\citenamefont{Henrich, Stratford,
  Marenduzzo, and Cates}}]{henrich2010ordering}
\bibinfo{author}{\bibfnamefont{O.}~\bibnamefont{Henrich}},
  \bibinfo{author}{\bibfnamefont{K.}~\bibnamefont{Stratford}},
  \bibinfo{author}{\bibfnamefont{D.}~\bibnamefont{Marenduzzo}},
  \bibnamefont{and} \bibinfo{author}{\bibfnamefont{M.~E.} \bibnamefont{Cates}},
  \bibinfo{journal}{Proc. Natl. Acad. Sci. USA} \textbf{\bibinfo{volume}{107}},
  \bibinfo{pages}{13212} (\bibinfo{year}{2010}).

\bibitem[{\citenamefont{Hagan and Chandler}(2006)}]{hagan2006dynamic}
\bibinfo{author}{\bibfnamefont{M.~F.} \bibnamefont{Hagan}} \bibnamefont{and}
  \bibinfo{author}{\bibfnamefont{D.}~\bibnamefont{Chandler}},
  \bibinfo{journal}{Biophys. J.} \textbf{\bibinfo{volume}{91}},
  \bibinfo{pages}{42} (\bibinfo{year}{2006}).

\bibitem[{\citenamefont{Whitesides and Boncheva}(2002)}]{whitesides2002bms}
\bibinfo{author}{\bibfnamefont{G.~M.} \bibnamefont{Whitesides}}
  \bibnamefont{and} \bibinfo{author}{\bibfnamefont{M.}~\bibnamefont{Boncheva}},
  \bibinfo{journal}{Proc. Nat. Acad. Sci.} \textbf{\bibinfo{volume}{99}},
  \bibinfo{pages}{4769} (\bibinfo{year}{2002}).

\bibitem[{\citenamefont{Jack et~al.}(2007)\citenamefont{Jack, Hagan, and
  Chandler}}]{break2}
\bibinfo{author}{\bibfnamefont{R.~L.} \bibnamefont{Jack}},
  \bibinfo{author}{\bibfnamefont{M.~F.} \bibnamefont{Hagan}}, \bibnamefont{and}
  \bibinfo{author}{\bibfnamefont{D.}~\bibnamefont{Chandler}},
  \bibinfo{journal}{Phys. Rev. E} \textbf{\bibinfo{volume}{76}},
  \bibinfo{pages}{21119} (\bibinfo{year}{2007}).

\bibitem[{\citenamefont{Rapaport}(2008)}]{new_rapaport}
\bibinfo{author}{\bibfnamefont{D.}~\bibnamefont{Rapaport}},
  \bibinfo{journal}{Phys. Rev. Lett.} \textbf{\bibinfo{volume}{101}},
  \bibinfo{pages}{186101} (\bibinfo{year}{2008}).

\bibitem[{\citenamefont{Marenduzzo et~al.}(2006)\citenamefont{Marenduzzo,
  Finan, and Cook}}]{marenduzzo2006depletion}
\bibinfo{author}{\bibfnamefont{D.}~\bibnamefont{Marenduzzo}},
  \bibinfo{author}{\bibfnamefont{K.}~\bibnamefont{Finan}}, \bibnamefont{and}
  \bibinfo{author}{\bibfnamefont{P.~R.} \bibnamefont{Cook}},
  \bibinfo{journal}{J. Cell Biol.} \textbf{\bibinfo{volume}{175}},
  \bibinfo{pages}{681} (\bibinfo{year}{2006}).

\bibitem[{\citenamefont{George et~al.}(1997)\citenamefont{George, Chiang, Guo,
  Arabshahi, Cai, and Wilson}}]{george19976}
\bibinfo{author}{\bibfnamefont{A.}~\bibnamefont{George}},
  \bibinfo{author}{\bibfnamefont{Y.}~\bibnamefont{Chiang}},
  \bibinfo{author}{\bibfnamefont{B.}~\bibnamefont{Guo}},
  \bibinfo{author}{\bibfnamefont{A.}~\bibnamefont{Arabshahi}},
  \bibinfo{author}{\bibfnamefont{Z.}~\bibnamefont{Cai}}, \bibnamefont{and}
  \bibinfo{author}{\bibfnamefont{W.~W.} \bibnamefont{Wilson}},
  \bibinfo{journal}{Methods in Enzymology} \textbf{\bibinfo{volume}{276}},
  \bibinfo{pages}{100} (\bibinfo{year}{1997}).

\end{thebibliography}
\end{document}